\documentstyle[aps,twocolumn]{revtex}
\input{psfig}
\begin{document}
\draft
\title{Shot Noise Current-Current Correlations in Multi-Terminal
Diffusive Conductors} 
\author{Ya.~M.~Blanter and M.~B\"uttiker}
\address{D\'epartement de Physique Th\'eorique, 
Universit\'e de Gen\`eve, CH-1211, Gen\`eve 4, Switzerland.}
\date{\today}
\maketitle 
\tighten

\begin{abstract}
We investigate the correlations in the current fluctuations
at different terminals of metallic diffusive conductors. We start from
scattering matrix expressions for the shot noise and use the
Fisher-Lee relation in combination with diagram technique to evaluate
the noise correlations. Of particular interest are exchange
(interference) effects analogous to the Hanbury Brown--Twiss effect in
optics. We find that the exchange effect exists in the ensemble
averaged current correlations. Depending on the geometry, it might
have the same magnitude as the mean square current fluctuations of the
shot noise. The approach which we use is first applied to present  
a novel derivation of the $1/3$-suppression of shot noise in a
two-terminal geometry, valid for an arbitrary relation between the length
and wire width. We find that in all geometries correlations are
insensitive to dephasing.  
\end{abstract}

\section{Introduction}

The shot noise in mesoscopic systems \cite{BdJ} continues to
attract the attention of both theorists and experimentalists. For
diffusive conductors, which are considered here, the
two-terminal shot noise is quite well studied. The spectacular
$1/3$-suppression of the shot noise with respect to the Poisson value,   
$$S(\omega=0) = \frac{1}{3} eGV$$
(here, as usual, $S(\omega)$ is the Fourier transform of the
current-current correlator, $S(t) = \langle \Delta I(t) \Delta I(0)
\rangle$, while $G$ and $V$ are the conductance of the wire and the
applied voltage, respectively; $\Delta I = I(t) - \langle I \rangle$),
was derived in three different ways: from the distribution of
transmission eigenvalues in a wire \cite{BB}, semi-classically from
the Langevin equation \cite{Nagaev}, and through a microscopic
calculation of local current densities \cite{ALY}. Later, Nazarov
\cite{Nazarov} claimed that this  $1/3$ suppression holds for
an arbitrary two-terminal geometry (not necessarily
quasi-one-dimensional). Subsequent to experiments by Liefrink et al
\cite{Liefrink}, which demonstrated shot-noise suppression close to
$1/3$ even for conductors much longer than the dephasing length, de
Jong and Beenakker \cite{semicl} provided a semi-classical discussion
which showed that the $1/3$-suppression is insensitive to
dephasing. More recent experiments by Steinbach et al \cite{Steinbach}
demonstrated the transition from the $1/3$-suppression regime in wires
short compared to an inelastic length through an interaction-dominated
regime \cite{Nagaev2,Kozub} to a regime where shot noise is suppressed
by inelastic scattering \cite{BB,Shimizu,Landauer}. A macroscopic
metal exhibits no shot noise \cite{Liu}. 

Here we investigate the shot noise in mesoscopic diffusive conductors
in a multi-terminal geometry. Primarily, we focus on the
interference experiment \cite{B92}, analogous to the experiment of
Hanbury Brown and Twiss in optics \cite{optics}. Namely, we consider
conductor, connected to four reservoirs $\alpha$, $\beta$, $\gamma$,
and $\delta$ at equilibrium (Fig.~1), and discuss three types of 
experiments. In the experiment A current is incident from the probe
$\beta$, i.e. $\mu_{\alpha} = \mu_{\gamma} = \mu_{\delta}$;
$\mu_{\beta} - \mu_{\alpha} = eV$, $\mu_{\lambda}$ being the chemical
potential of electrons in the reservoir ${\lambda}$.  In the
experiment B current is incident from the probe $\delta$:
$\mu_{\alpha} = \mu_{\beta} = \mu_{\gamma}$; $\mu_{\delta} -
\mu_{\alpha} = eV$. Finally, in the experiment C current is incident
from both probes $\beta$ and $\delta$: $\mu_{\alpha} = \mu_{\gamma}$;
$\mu_{\beta} = \mu_{\delta}$; $\mu_{\beta} - \mu_{\alpha} = eV$. 
The current correlation in probes $\alpha$ and $\gamma$ is measured in
all the experiments, $S_j(t) = - \langle \Delta I_{\alpha} (t) \Delta
I_{\gamma} (0) \rangle$, $j = A,B,C$. 
  
The general analysis of Ref. \cite{B92} allows to express these
quantities in terms of scattering matrices $\hat s^{\lambda \nu}$,
with indices $\lambda$ and $\nu$ labeling the probes. Thus, for zero
frequency and temperature \cite{foot1} one obtains
\begin{eqnarray} \label{basic}
\left\{ \begin{array}{l} S_A \\ S_B \\ S_C \\ \end{array} \right\}
 = \frac{e^2}{\pi}e\vert V \vert
\left\{ \begin{array}{c} \Xi_1\\  \Xi_2\\ \Xi_1 + \Xi_2 + \Xi_3 +
 \Xi_4 \end{array} \right\},
\end{eqnarray}
with quantities $\Xi_i$ defined as follows,
\begin{eqnarray} \label{HBT}
\Xi_1 & = & \mbox{Tr}(s^{+\alpha\beta} s^{\alpha\beta}
s^{+\gamma\beta} s^{\gamma\beta}); 
\nonumber \\   
\Xi_2 & = & \mbox{Tr}(s^{+\alpha\delta} s^{\alpha\delta}
s^{+\gamma\delta} s^{\gamma\delta}); 
\nonumber \\
\Xi_3 & = & \mbox{Tr}(s^{+\alpha\beta} s^{\alpha\delta}
s^{+\gamma\delta} s^{\gamma\beta}); 
\nonumber \\
\Xi_4 & = & \mbox{Tr}(s^{+\alpha\delta} s^{\alpha\beta}
s^{+\gamma\beta} s^{\gamma\delta}), 
\end{eqnarray}
the scattering matrices are evaluated at the Fermi surface, and the trace
is taken with respect to channel indices. 

Thus, $S_C \ne S_A + S_B$: experiments A and B are not additive due to
the interference terms $\Xi_3$ and $\Xi_4$. It was shown in
Ref. \cite{B92} that these terms have different signs for fermions and
bosons; hence we will call them exchange terms. We define an exchange
contribution as  
$$\Delta S  = S_C - S_A - S_B.$$

It follows from the unitarity of matrices $s^{\lambda\nu}$
that the quantities $\Xi_1$ and $\Xi_2$, which represent the classical
result, are positively defined \cite{foot2}. At the same time, traces
$\Xi_3$ and $\Xi_4$ can have either sign. This means that
exchange-interference may either suppress or enhance the classical
value.   

In a disordered system all these quantities should be averaged over
impurity configuration. Naively, one might think that due to the phases
contained in the quantities $\Xi_3$ and $\Xi_4$ these will average to
zero, and thus the average of the exchange term $\langle \Delta S
\rangle$ vanishes (here angular brackets are used to indicate the
disorder average). Below we explicitly calculate disorder-averaged
correlation functions $S_j$, and demonstrate that it is not the
case. The average exchange correlator $\langle \Delta S \rangle$
generally has a nonzero value. An analysis of the exchange correlator
for chaotic cavities, reported elsewhere \cite{Langen}, leads to a
similar conclusion. 

\begin{figure}
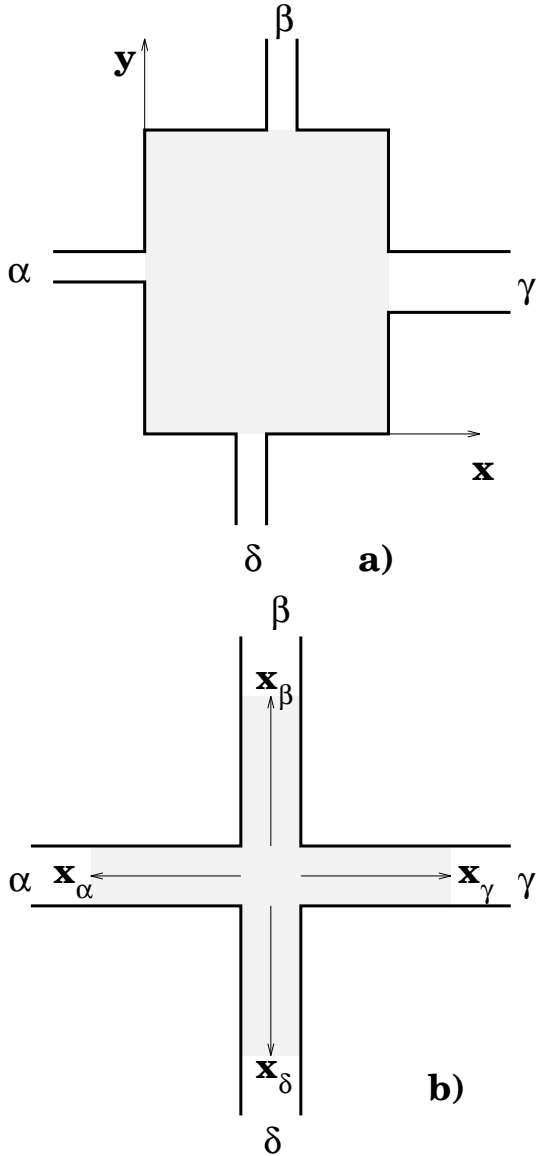

\centerline{\psfig{figure=nois1a.eps,width=7.cm}}
\centerline{\psfig{figure=nois1b.eps,width=7.cm}}
\vspace{0.3cm}
\caption{Four-terminal conductors; the disordered area is shaded.}
\label{picture}
\end{figure}

The paper is organized as follows. First, we investigate 
disorder averages of scattering matrices starting from
Eqs. (\ref{HBT}) and using the Fisher-Lee relation which connects
scattering matrices and Green's functions. We then use diagram
techniques developed for disordered systems to find the ensemble
averages. As a simple check of the method developed, we give a novel
derivation of the $1/3$-suppression of the two-terminal shot noise for
an arbitrary (not necessarily quasi-one-dimensional) geometry, thus
confirming the result by Nazarov \cite{Nazarov}. Then we turn to
exchange-interference experiments and consider the two particular
four-terminal geometries, shown in Fig.~1. We demonstrate that the
geometry of Fig.~1a implies a negative exchange correlation, with the
quantity $\Delta S$ being of the same order of magnitude as
correlators $S_A$ and $S_B$ themselves. In contrast, the cross
geometry of Fig.~1b shows a strong suppression of exchange effects,
and gives a positive sign of the latter, provided the motion through
the center of a cross is ballistic. Otherwise the exchange effect is
governed by the scattering inside the cross center only.   

In the calculations below we disregard electron-electron
interaction. The latter is known not to produce an essential effect on
two-terminal shot noise \cite{Nagaev2,Kozub} provided the wire is
short in comparison with the inelastic scattering length. We will show
that the origin for this is that in the ensemble averaged quantities
the effect is local and electron trajectories enclosing a large area
are suppressed. This explains why the shot noise is not sensitive to
dephasing. Hence, we believe that electron-electron interactions are
not important for the exchange effects in shot noise. Note, however,
that non-linear noise is affected by interactions, as was
shown recently \cite{Stern}. Interactions are also expected to affect
the frequency dependence of the shot noise power.  

\section{General formalism and two-terminal shot noise}

We consider a disordered two-dimensional system, connected to
reservoirs by ideal leads. Transverse motion of electrons in each lead
is quantized, and we assume that all leads are wide, i.e. the number
of transverse channels at the Fermi surface in the lead $\lambda$ is
large, $N_{\lambda} = p_FW_{\lambda} \gg 1$. Here $p_F$ is the Fermi 
momentum, while $W_{\lambda}$ is the width of the lead.  

General relations \cite{FLee,Stone,BS} allow one to express
scattering matrices for an arbitrary geometry through retarded and
advanced Green's functions of the system. The standard procedure
\cite{BS} is as follows. One chooses arbitrary cross-sections of the
leads $C_{\lambda}$, and introduces local coordinates related to these
cross-sections (Fig.~2). Since nothing depends on the choice of these
cross-sections, it is convenient to choose them as boundary between
disordered region and leads. One obtains
\begin{figure}
\centerline{\psfig{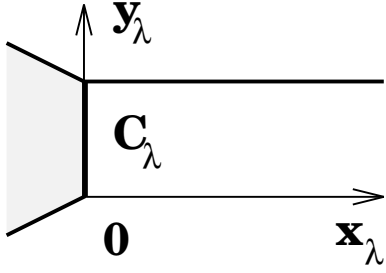}}
\vspace{0.3cm}
\caption{Contact of a disordered region (shaded) with an ideal lead
$\lambda$.} 
\label{junc}
\end{figure}

\begin{eqnarray} \label{FL}
s^{\lambda\nu}_{mn} (E) & = & -\frac{i}{4M^2(v_mv_n)^{1/2}}
\int_{C_{\lambda}} dy_{\lambda} 
\int_{C_{\nu}} dy_{\nu} G_E^R
(\bbox{r}_{\lambda},\bbox{r}_{\nu})\nonumber \\
& \times &  (\bbox{D}_{\lambda} \hat{\bbox{n}}_{\lambda}) (\bbox{D}_{\nu}
\hat{\bbox{n}}_{\nu}) \exp (-ik_mx_{\lambda} - ik_n x_{\nu})
\nonumber \\
& \times & \chi_m(y_{\lambda}) \chi_n(y_{\nu})
\end{eqnarray}
and 
\begin{eqnarray} \label{FL1}
s^{+\lambda\nu}_{nm} (E) & = & \frac{i}{4M^2(v_mv_n)^{1/2}}
\int_{C_{\lambda}} dy_{\lambda} \int_{C_{\nu}} dy_{\nu} G_E^A
(\bbox{r}_{\nu},\bbox{r}_{\lambda})\nonumber \\
& \times &  (\bbox{D}_{\lambda} \hat{\bbox{n}}_{\lambda}) (\bbox{D}_{\nu}
\hat{\bbox{n}}_{\nu}) \exp (ik_mx_{\lambda} + ik_n x_{\nu})
\nonumber \\
& \times & \chi_m(y_{\lambda}) \chi_n(y_{\nu}).
\end{eqnarray}
Here $v_m = k_m/M$, $M$ being the effective electron mass. The
longitudinal wavevectors in the lead $\lambda$ are
$$k_m = \left [p_F^2 - \left( \pi m/W_{\lambda} \right)^2
\right]^{1/2},$$ 
and those in lead $\nu$ are denoted by $k_n$. Furthermore,
$\hat{\bbox{n}}_{\lambda}$ is the unit vector in the direction
$x_{\lambda}$, while $\chi_m$ and $\chi_n$ are wavefunctions of
transverse motion in the leads $\lambda$ and $\nu$ respectively; for
simplicity we choose them to be real. Finally, $\bbox{D}$ denotes
a double-sided derivative,  
$$f \bbox{D} g = f \nabla g - g \nabla f.$$ 

In principle, Eqs. (\ref{FL}) and (\ref{FL1}) allow one to average
arbitrary combinations of scattering matrices over disorder, using
the standard diagram technique \cite{AGD}. It seems that the approach
outlined here has not so far been used for the (analytical)
calculation of any physical properties. However, it is rather close to
the Hamiltonian approach, employed extensively for the calculation of
conductance and conductance fluctuations
\cite{IWZ,Altland,IM,MG,MMZ}. Below we demonstrate that our formalism
reproduces the $1/3$-suppression of two-terminal shot noise; in
particular, as a simplest check, we also reproduce the Drude formula
for conductance. 

The rest of the Section is devoted to the two-terminal geometry --- a
diffusive wire of the length $L$ and width $W$, connected to two ideal
leads $\alpha$ and $\beta$; $L,W \gg l$, with $l$ being the mean free
path. For a moment we assume also $L \gg W$, a restriction, which
eventually will be lifted.  We introduce an axis $\hat x$ directed
along the wire, $0 \le x \le L$, and an axis $\hat y$ directed across
the wire. The general expressions (\ref{FL}), (\ref{FL1}) can be
rewritten as 
\begin{eqnarray} \label{FL2}
& & s^{\alpha\beta}_{mn} (E) = \frac{i}{4M(k_mk_n)^{1/2}}
\int_{C_{\alpha}} dy_1 \chi_m(y_1) \int_{C_{\beta}} dy_2 \chi_n(y_2)
\nonumber \\ & & \times [-\partial_{x_1} + ik_m][-\partial_{x_2} -
ik_n] G^R_E(\bbox{r}_1,\bbox{r}_2) \vert_{x_2=L}^{x_1=0,} 
\end{eqnarray}
and 
\begin{eqnarray} \label{FL21}
s^{+\alpha\beta}_{mn} (E) & = & -\frac{i}{4M(k_mk_n)^{1/2}}
\int_{C_{\alpha}} dy_1 \chi_n(y_1) \int_{C_{\beta}} dy_2  
\nonumber \\  
& \times & \chi_m(y_2) [-\partial_{x_1} - ik_n][-\partial_{x_2} +
ik_m] \nonumber \\
& \times & G^A_E(\bbox{r}_2,\bbox{r}_1) \vert_{x_2=L}^{x_1=0,}\ . 
\end{eqnarray}

Two-terminal shot noise power $S \equiv S(\omega = 0)$ can be
conveniently expressed through scattering matrices evaluated at the
Fermi level \cite{B90,Khlus},
\begin{equation} \label{2TSN}
S = \frac{e^2}{2\pi} eV \langle \mbox{Tr} \left[
s^{+\alpha\beta}s^{\alpha\beta} \right] 
- \mbox{Tr} \left[
s^{+\alpha\beta}s^{\alpha\beta}s^{+\alpha\beta}s^{\alpha\beta} \right]
\rangle.  
\end{equation}
Note that the first trace on the rhs is related to the conductance,
$$G = \frac{e^2}{2\pi} \langle \mbox{Tr} \left[
s^{+\alpha\beta}s^{\alpha\beta} \right] \rangle.$$
It is convenient to calculate both traces separately.

\subsection{Evaluation of $\langle \mbox{Tr} \left[
s^{+\alpha\beta}s^{\alpha\beta} \right] \rangle.$} 

Using Eqs.(\ref{FL2}) and (\ref{FL21}), we find for the conductance 
\begin{eqnarray} \label{g1}
& & g \equiv \langle \mbox{Tr} \left[ s^{+\alpha\beta}s^{\alpha\beta}
\right] 
\rangle = \frac{1}{(4M)^2} \sum_{mn} \frac{1}{k_mk_n} \nonumber \\
& & \times  \int_{C_{\alpha}} dy_2
dy_3 \chi_n(y_2) \chi_n(y_3) \int_{C_{\beta}} dy_1 dy_4 \chi_m(y_1)
\chi_m(y_4) \nonumber \\
& & \times \left[ ik_m - \partial_{x_1} \right] \left[ -ik_n -
\partial_{x_2} \right] \left[ ik_n - \partial_{x_3} \right] \left[
-ik_m - \partial_{x_4} \right] \nonumber \\
& & \times \langle G^A(\bbox{r}_1,\bbox{r}_2)
G^R(\bbox{r}_3,\bbox{r}_4) \rangle \vert^{x_1=x_4=L,}_{x_2=x_3=0}\ , 
\end{eqnarray}
where the Green's functions are taken at the Fermi energy. Since the
averaged Green's functions decay on scales of the mean free path, the
average product of two Green's functions, each of them taken in remote
points, is only due to the diffusion (see e.g. \cite{AA}):
\begin{eqnarray} \label{diff1}
& & \langle G^A (\bbox{r}_1,\bbox{r}_2) G^R (\bbox{r}_3,\bbox{r}_4)
\rangle = \int d\bbox{r}_a d\bbox{r}_b \langle
G^A(\bbox{r}_1,\bbox{r}_a) \rangle
\nonumber \\
& & \times \langle G^A(\bbox{r}_b,\bbox{r}_2) \rangle \langle G^R
(\bbox{r}_3,\bbox{r}_b) \rangle \langle G^R 
(\bbox{r}_a,\bbox{r}_4) \rangle P(\bbox{r}_a, \bbox{r}_b).
\end{eqnarray}
The diffusion propagator $P(\bbox{r}, \bbox{r'})$ is a solution of the
equation
\begin{equation} \label{diff2}
-D\nabla^2_{\bbox{r}} P(\bbox{r}, \bbox{r'}) = (2\pi\nu\tau^2)^{-1}
\delta(\bbox{r}-\bbox{r'}) 
\end{equation}
with appropriate boundary conditions ($P=0$ at the contact to the ideal
leads; $\bbox{n} \nabla P = 0$ at the walls). Here $\nu = M/2\pi$, $D
= v_Fl/2$ and $\tau$ are density of states, diffusion coefficient, and
elastic lifetime, respectively. Under the assumption $L \gg W$ the
diffusion can be considered to be one-dimensional, and the diffusion
propagator does not depend on $y$,
\begin{eqnarray} \label{diff3}
P(x, x') = (M\tau^2DWL)^{-1} \left\{ \begin{array}{lr} 
x(L-x'), & \ x < x'\\ 
x'(L-x), & \ x > x'
\end{array} \right. .  
\end{eqnarray} 
\begin{figure}
\centerline{\psfig{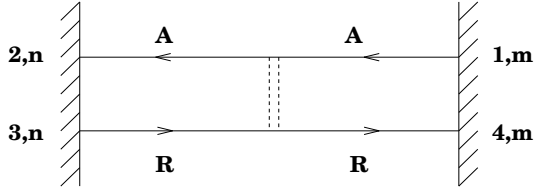}}
\vspace{0.3cm}
\caption{Diagram for the conductance. The double dashed line
represents the diffusion propagator. The position and the transverse
channel number of the points on the surfaces $C_{\alpha}$ $(x=0)$ and
$C_{\beta}$ $(x=L)$ are shown. For example, the transverse wavefunction
$\chi_m$ is taken at the point $y_1$.}   
\label{gdiag}
\end{figure}
Now we insert Eq. (\ref{diff1}) into Eq. (\ref{g1}). The diagram for 
$g$ is shown in Fig.~3. One can approximate the short-ranged Green's
functions as follows,
\begin{eqnarray} \label{green1}
\langle G^R (\bbox{r},\bbox{r'}) \rangle & = & -\frac{iM}{p_F}
\exp \left[ \left (ip_F - \frac{1}{2l} \right) \vert x - x' \vert
\right] \nonumber \\
& \times & \delta(y - y'). 
\end{eqnarray}
Then, integrating over transverse coordinates, we obtain 
\begin{eqnarray} \label{g2}
g & = & \frac{M}{16D\tau^2LW} \left[ \sum_m \frac{1}{k_m} \left(1 +
\frac{k_m}{p_F}\right)^2 \right]^2 \int_0^L dx_a dx_b \nonumber \\
& \times & \exp[-x_a/l] \exp[-(L-x_b)/l] x_a (L-x_b).
\end{eqnarray}
Taking into account that
$$\sum_m \frac{1}{k_m} \left(1 + \frac{k_m}{p_F}\right)^2 = 2W,$$
we obtain
\begin{equation} \label{fing}
g = \frac{l}{2L} p_FW.
\end{equation}
Multiplied by $e^2/2\pi$, Eq. (\ref{fing}) gives the Drude formula, as it
should be. 

\subsection{Evaluation of $\langle \mbox{Tr} \left[
s^{+\alpha\beta}s^{\alpha\beta}s^{+\alpha\beta}s^{\alpha\beta} \right]
\rangle.$} 

The trace of a product of four scattering matrices can be written as 
\begin{eqnarray} \label{t1}
& & t \equiv \langle \mbox{Tr} \left[
s^{+\alpha\beta}s^{\alpha\beta}s^{+\alpha\beta}s^{\alpha\beta} \right]
\rangle = \frac{1}{(4M)^4} \sum_{klmn} \frac{1}{k_kk_lk_mk_n}
\nonumber \\
& & \times \int_{C_{\alpha}} dy_2 dy_3 dy_6 dy_7 \chi_l(y_2)
\chi_l(y_3) \chi_n(y_6) \chi_n(y_7) \nonumber \\ 
& & \times \int_{C_{\beta}} dy_1 dy_4 dy_5 dy_8 \chi_k(y_1)
\chi_m(y_4) \chi_m(y_5) \chi_k(y_8) \nonumber \\
& & \times \left[ ik_k - \partial_{x_1} \right] \left[ -ik_l -
\partial_{x_2} \right] \left[ ik_l - \partial_{x_3} \right] \left[ -ik_m -
\partial_{x_4} \right] \nonumber \\
& & \times \left[ ik_m - \partial_{x_5} \right] \left[ -ik_n -
\partial_{x_6} \right] \left[ ik_n - \partial_{x_7} \right] \left[ ik_k -
\partial_{x_8} \right] \nonumber \\
& & \times \langle G^A(\bbox{r}_1,\bbox{r}_2)
G^R(\bbox{r}_3,\bbox{r}_4) G^A(\bbox{r}_5,\bbox{r}_6) \nonumber \\
& & \times G^R(\bbox{r}_7; \bbox{r}_8) \rangle
\vert^{x_1=x_4=x_5=x_8=L,}_{x_2=x_3=x_6=x_7=0}\ .
\end{eqnarray}
Employing Eq. (\ref{diff1}) again, we find the diagrams shown in
Fig.~4. We omitted all diagrams containing a single electron line
connecting two different leads, since these are exponentially small;
the diagrams (a) and (e) contain also counterparts, similar to (c) and
(d).   
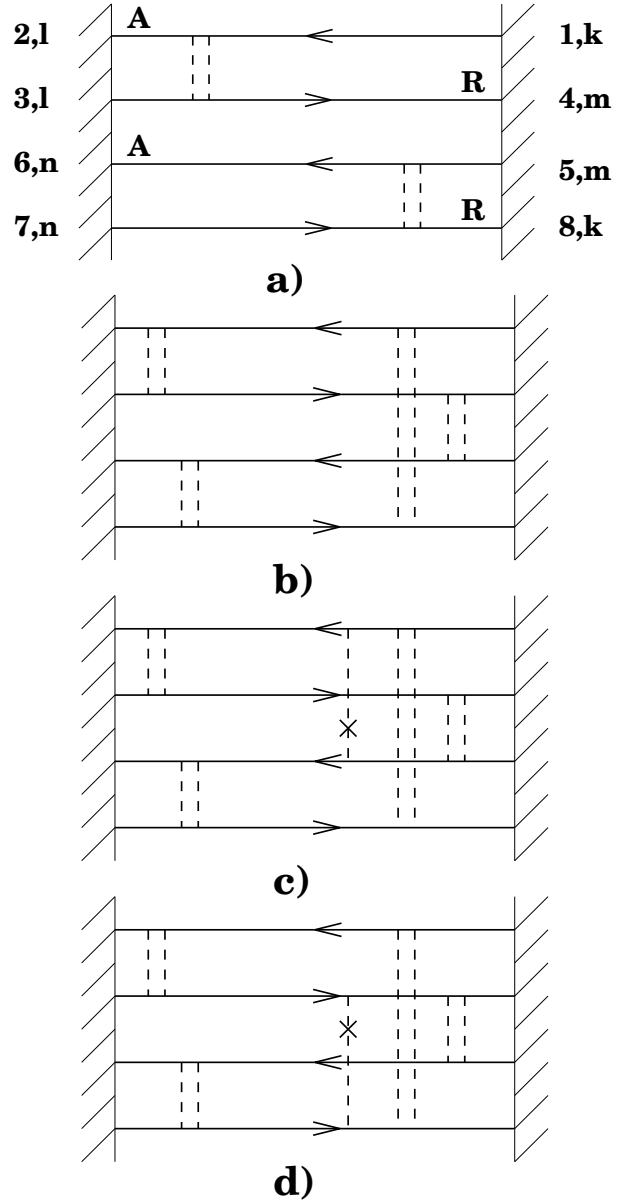
\begin{figure}
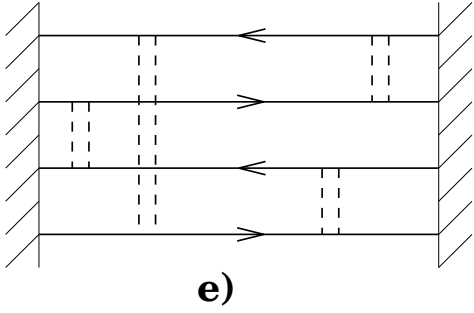

\centerline{\psfig{figure=nois4a.eps,width=8cm}}
\centerline{\psfig{figure=nois4b.eps,width=6.2cm}}
\centerline{\psfig{figure=nois4c.eps,width=6.2cm}}
\centerline{\psfig{figure=nois4d.eps,width=6.2cm}}
\centerline{\psfig{figure=nois4e.eps,width=6.2cm}}
\vspace{0.3cm}
\caption{Diagrams for the quantity $t$ in the same notations as in
Fig.~3. The single dashed line with a cross represents impurity
scattering.}    
\label{gdiag1}
\end{figure}
The diagrams (b), (c) and (d) turn out to give the leading
contribution, whereas others carry small factors. Thus, for diagram
(a) points $y_1$ and $y_4$ should lie not further apart than a mean
free path, which due to the orthogonality of transverse wavefunctions
implies $k=m$. Therefore, the contribution of this diagram is
suppressed by a factor $(p_FW)^{-1} \ll 1$. The diagram (e), which is
topologically equivalent to (b), is suppressed as
$(p_FW)^{-3}$. Taking into account the explicit form (\ref{diff3}) for
the diffusion propagator, and integrating over coordinates $y_i$ and
over one of two pair of coordinates in the diffusion propagators
(those lying close to one of the ends of the wire), we arrive at the
expression   
\begin{eqnarray} \label{t2}
t & = & \frac{l^8}{2(4D\tau^2WL)^4} \left[ \sum_m \frac{1}{k_m} \left(
1+ \frac{k_m}{p_F} \right)^2 \right]^4 \int d\bbox{r}_a d\bbox{r}_b
d\bbox{r}_c \nonumber \\ 
& \times & d\bbox{r}_d (L-x_a) (L - x_c) x_b x_d F(\bbox{r}_a,
\bbox{r}_b, \bbox{r}_c, \bbox{r}_d). 
\end{eqnarray}
Here $F$ is the Hikami box \cite{Hikami}. It is short-ranged (all
points $\bbox{r}_a$, $\bbox{r}_b$, $\bbox{r}_c$, and $\bbox{r}_d$
should be close to each other), and in the Fourier-space has the form 
\begin{eqnarray} \label{Hik1}
& & F(\bbox{q}_a, \bbox{q}_b, \bbox{q}_c, \bbox{q}_d) =
-M\tau^5v_F^2(2\pi)^2 \delta(\bbox{q}_a +\bbox{q}_b + \bbox{q}_c +
\bbox{q}_d) \nonumber \\
& & \times \left[ 2(\bbox{q}_a \bbox{q}_c + \bbox{q}_b \bbox{q}_d) +
(\bbox{q}_a + \bbox{q}_c)(\bbox{q}_b + \bbox{q}_d) \right].  
\end{eqnarray}
Integration of the Hikami box over the cross-section of the wire
yields 
\begin{eqnarray} \label{Hik2}
& & \int F(\bbox{r}_a, \bbox{r}_b, \bbox{r}_c, \bbox{r}_d) dy_a dy_b
dy_c dy_d = M\tau^5v_F^2W \left[ 2\partial_{x_a}\partial_{x_c} \right.
\nonumber \\
& & \left. + 2\partial_{x_b}\partial_{x_d} +
\partial_{x_a}\partial_{x_b} + \partial_{x_a}\partial_{x_d} +
\partial_{x_b}\partial_{x_c} + \partial_{x_c}\partial_{x_d} \right]
\nonumber \\
& & \times \delta(x_a-x_b) \delta(x_a-x_c) \delta(x_a - x_d).  
\end{eqnarray} 
Inserting Eq. (\ref{Hik2}) into Eq. (\ref{t2}) and performing the
remaining integrations, we obtain
\begin{equation} \label{t3}
t = \frac{l}{3L} p_FW = 2g/3
\end{equation}
which immediately gives the $1/3$-shot noise suppression. 

\subsection{Universality} 

Now we lift the requirement $L \gg W$, but still consider a diffusive
system, $W, L \gg l$. The result (\ref{fing}) for $g = \langle
\mbox{Tr} [s^{+\alpha\beta} s^{\alpha\beta}] \rangle$ is equivalent,
in fact, to the Drude formula, and is therefore valid for an 
arbitrary relation between $W$ and $L$. In the derivation of $t =
\langle \mbox{Tr} [s^{+\alpha\beta} s^{\alpha\beta} s^{+\alpha\beta}
s^{\alpha\beta}] \rangle$ we should now take into account that the
diffusion is not one-dimensional any more, and write the diffusion
propagator in the form 
\begin{equation} \label{un1}
P(\bbox{r}, \bbox{r}') = \frac{1}{M\tau^2D} \sum_{\bbox{q}}
\frac{1}{q^2} \phi_{\bbox q} (\bbox{r}) \phi_{\bbox q} (\bbox{r}'),
\end{equation}  
instead of Eq. (\ref{diff3}). Here $\phi_{\bbox{q}} (\bbox{r})$ and
$-q^2$ are eigenfunctions and eigenvalues of the Laplace operator with
appropriate boundary conditions. In our particular geometry one
obtains 
$$\bbox{q} = (\frac{\pi}{L} n_x, \frac{\pi}{W} n_y),$$  
with integers $n_x > 0$ and $n_y \ge 0$. It is
easy to see that the integration over $y_1$ and $y_8$ in the diagrams
of Fig.~4 (b),(c),(d) places a constraint on the wavevector $n_{1y}$
of the diffusion propagator connecting these two points, $n_{1y} = 2k$
(unless $n_{1y} = 0$). In the same way, the other integrations over
$y_i$ imply other constraints, which due to the $\delta$-function in
the expression for the Hikami box (\ref{Hik1}) yield a constraint on
the channel indices $k,l,m,n$. Therefore all terms with non-zero
transverse harmonics are small as $(p_FW)^{-1}$. Up to terms
proportional to this small parameter the result (\ref{t3}) is
exact. Thus, the $1/3$-shot noise suppression is, indeed, universal,
and does not depend on the ratio $W/L$, provided the system is
diffusive, in accordance with the conclusion of Ref. \cite{Nazarov}. 

To conclude this section, we compare the method used above with other
derivations of the $1/3$-shot noise suppression
\cite{BB,Nagaev,ALY}. As is well known, there exist two principally
different methods of calculating conductance. One can first evaluate
conductivity (which is a local quantity), starting from the Kubo
formula, and then, after integration over a cross-section one obtains
the conductance. Alternatively, one can calculate conductance
directly, starting from the Landauer formula. (In fact, our derivation
of the quantity $g$ given above is of this kind). Both derivations are
equivalent, although at intermediate stages they have not much in
common.   

A similar situation happens in the calculation of shot noise. On one
hand, one can calculate the microscopic correlator of currents, and
upon integration over a cross-section obtains the shot noise power. The
derivation of Altshuler, Levitov and Yakovets \cite{ALY} is exactly of
this type \cite{KhL}. It can be generalized to an arbitrary geometry,
and, in principle, can be used for a broad class of problems. The
local current correlator contains more information than is necessary
for the calculation of the shot noise power. The method 
of Nagaev \cite{Nagaev} and de Jong and Beenakker \cite{semicl}, who
employ the quantum Langevin equation, is 
somewhat similar, although the equivalence between these two
approaches is not evident. The generalization of the latter approach
for a multi-terminal geometry does not seem to be quite obvious.  

The derivation of Beenakker and one of the authors \cite{BB}, as well
as the present method, belong to another, scattering (or Landauer)
type of approaches. Ref. \cite{BB} derives the shot-noise power with
the use of the distribution of transmission eigenvalues of diffusive
wire. This proof seems to be the most elegant. However, one should not
forget that the distribution of transmission eigenvalues itself is
derived by sophisticated methods such as the DMPK equation
\cite{DMPK}. Although Nazarov \cite{Nazarov} succeeded in extending
this derivation to the case of an arbitrary two-terminal geometry,
most probably it can not be generalized to multi-terminal case: for
conductors with four (or more) probes the shot noise is not expressed
through eigenvalues of the scattering matrix $s^+s$. The
derivation given in this paper is more general, and self-contained; it
does not require the distribution of transmission eigenvalues, and
hence allows a generalization to an arbitrary scattering geometry. We
repeat that it differs from the  microscopic calculation of
ref. \cite{ALY}. Certainly, although these two approaches can not be
easily compared at intermediate stages, in the end they should lead to
identical results for arbitrary scattering geometries.

\section{Multi-terminal shot noise}

Now we turn to the exchange--interference experiment described in the
Introduction. We consider a four-terminal geometry (examples are shown
in Fig.~1; for convenience, we still use the coordinates of Fig.~2),
and calculate the current correlation for experiments A, B, and C.  

The quantity $\Xi_1 = \mbox{Tr}(s^{+\alpha\beta} s^{\alpha\beta}
s^{+\gamma\beta} s^{\gamma\beta})$ is 
determined by diagrams of Fig.~4, where now points $y_1,y_4,y_5,y_8$
belong to the contact with lead $\beta$; points $y_2,y_3$ and
$y_6,y_7$ belong to the contacts with leads $\alpha$ and $\gamma$,
respectively. Therefore the diagram (e) is exponentially
small, while the diagram (a) is suppressed in the parameter
$(p_FW_{\beta})^{-1}$. Hence, the quantity $\Xi_1$ is given by the
same diagrams (b),(c), and (d), as the quantity $t$. Using
Eq. (\ref{green1}) and integrating over the cross-section of the
leads, we obtain (Fig.~5)  
\begin{eqnarray} \label{xi1g}
& & \Xi_1 = \frac{1}{2} \left( \frac{M}{4} \right)^4 \sum_{klmn}
\frac{1}{k_kk_lk_mk_n} \left( 1 + \frac{k_k}{p_F} \right)^2 \left( 1 +
\frac{k_l}{p_F} \right)^2 \nonumber \\
& & \times \left( 1 + \frac{k_m}{p_F} \right)^2 \left(
1 + \frac{k_n}{p_F} \right)^2 \int_{C_{\alpha}} dy_f \chi_l^2 (y_f)
\int_{C_{\beta}} dy_a dy_c \nonumber \\
& & \times \chi_k^2 (y_a) \chi_m^2 (y_c) \int_{C_{\gamma}} dy_h
\chi_n^2 (y_h) \int_{-\infty}^0 dx_a dx_c dx_f dx_h \nonumber \\
& & \times \exp((x_a + x_c + x_f + x_h)/l) \int d\bbox{r}_a \dots
d\bbox{r}_h P(\bbox{r}_a, \bbox{r}_b) \nonumber \\
& & \times P(\bbox{r}_c, \bbox{r}_d) P(\bbox{r}_e,
\bbox{r}_f) P(\bbox{r}_g, \bbox{r}_h) F(\bbox{r}_b, \bbox{r}_e,
\bbox{r}_d, \bbox{r}_g).  
\end{eqnarray}
Here the points $\bbox{r}_a, \bbox{r}_c, \bbox{r}_f, \bbox{r}_h$ are
given in the coordinates of the contacts $\beta$, $\beta$, $\alpha$,
$\gamma$, respectively.
\begin{figure}
\centerline{\psfig{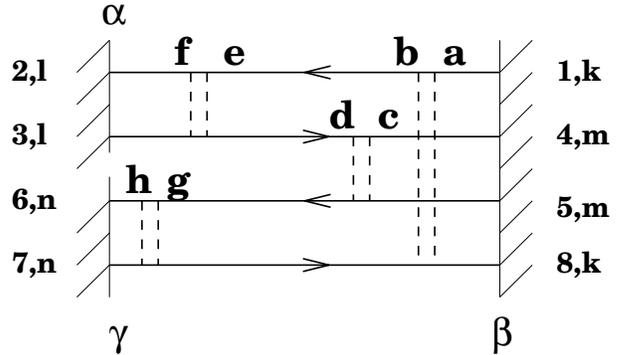}}
\vspace{0.3cm}
\caption{Typical diagram for the quantity $\Xi_1$.}   
\label{gdiag5}
\end{figure}

In the same way, for the quantity $\Xi_3$ one obtains 
\begin{eqnarray} \label{xi3g}
& & \Xi_3 = \frac{1}{2} \left( \frac{M}{4} \right)^4 \sum_{klmn}
\frac{1}{k_kk_lk_mk_n} \left( 1 + \frac{k_k}{p_F} \right)^2 \left( 1 +
\frac{k_l}{p_F} \right)^2 \nonumber \\
& & \times \left( 1 + \frac{k_m}{p_F} \right)^2 \left(
1 + \frac{k_n}{p_F} \right)^2 \int_{C_{\alpha}} dy_f \chi_l^2 (y_f)
\int_{C_{\beta}} dy_a \chi_k^2 (y_a) \nonumber \\
& & \times \int_{C_{\gamma}} dy_h
\chi_n^2 (y_h) \int_{C_{\delta}} dy_c \chi_m^2 (y_c)
\int_{-\infty}^0 dx_a dx_c dx_f dx_h \nonumber \\
& & \times \exp((x_a + x_c + x_f + x_h)/l) \int d\bbox{r}_a \dots
d\bbox{r}_h P(\bbox{r}_a, \bbox{r}_b) \nonumber \\
& & \times P(\bbox{r}_c, \bbox{r}_d) P(\bbox{r}_e,
\bbox{r}_f) P(\bbox{r}_g, \bbox{r}_h) F(\bbox{r}_b, \bbox{r}_e,
\bbox{r}_d, \bbox{r}_g),  
\end{eqnarray}
and the points $\bbox{r}_a, \bbox{r}_c, \bbox{r}_f, \bbox{r}_h$ are
given in the coordinates of the contacts $\beta$, $\delta$, $\alpha$, 
$\gamma$, respectively. Expressions for the quantities $\Xi_2$ and
$\Xi_4$ can be obtained from Eqs. (\ref{xi1g}) and (\ref{xi3g}),
respectively, by interchanging $\beta \leftrightarrow \delta$.

Expressions (\ref{xi1g}) and (\ref{xi3g}) are valid for an arbitrary
four-terminal geometry and can be used for numerical calculations. It
is important that not only traces $\Xi_1$ and $\Xi_2$, as one could
expect, but also quantities $\Xi_3$ and $\Xi_4$ are phase
insensitive. Indeed, the electron motion which Eqs. (\ref{xi1g}) and
(\ref{xi3g}) imply is just the diffusion between different leads. No
closed paths are formed, except for ballistic motion due to the
scattering described by the Hikami box somewhere in the middle of the
sample. Since the size of this loop is very small, of order of the
mean free path, dephasing is not expected to have an effect on the
exchange noise. Certainly, some effects similar to weak localization
exist, however, as for conductance \cite{AA}, they are relatively weak
(as $(p_Fl)^{-1}$) in comparison with the main effect. We do not
discuss these effects here. 

To make further progress we have to solve the diffusion equation in a
given geometry with appropriate boundary conditions. We turn now to
the two different geometries, shown in Fig.~1.

\subsection{Box geometry}

First, we consider the geometry of Fig.~1a. We assume all leads to be
wide, $W_{\lambda} \gg l$. Then points $\bbox{r}_a$, $\bbox{r}_c$,
$\bbox{r}_f$, and $\bbox{r}_h$ are typically far from the lead's
boundaries. This means that, for example, in the integral over $y_f$
one can replace the diffusion propagator, $P(\bbox{r}_e, \bbox{r}_f)$
by another function $\tilde P(\bbox {r}_e, \bbox{r}_f)$, which is also
a solution to the diffusion equation, but with another boundary
conditions, appropriate for an open surface,
$$\tilde P(\bbox{r}, \bbox{r'}) \vert_{x=0} = 0.$$
We do not need to specify boundary conditions for $\tilde P(\bbox
{r}_e, \bbox{r}_f)$ on the other boundaries, since the point
$\bbox{r}_e$ is typically in the middle of the sample. Consequently,
we may substitute for all ``true'' diffusion propagators $P$ the
functions $\tilde P$, the solution with $\tilde P = 0$ everywhere on
the boundary, as is appropriate for an open system. The solution
$\tilde P$ is
\begin{eqnarray} \label{box0}
& & \tilde P(\bbox{r}, \bbox{r'}) = \frac{4}{M\tau^2 DL_xL_y}
\sum_{n_x,n_y=1}^{\infty} \frac{1}{\pi^2 n_x^2/L_x^2 + \pi^2
n_y^2/L_y^2} \nonumber \\ 
& & \times \sin \frac{\pi n_x x}{L_x} \sin \frac{\pi n_x x'}{L_x}
\sin \frac{\pi n_y y}{L_y} \sin \frac{\pi n_y y'}{L_y}. 
\end{eqnarray}
Furthermore, the functions $\tilde P$ vary
considerably on the scale of the size of a sample, $L_x$ and $L_y$. If
we assume $W_{\lambda} \ll L_x,L_y$, the function $\tilde P(\bbox
{r}_e, \bbox{r}_f)$ in the integral over $y_f$ may be taken
independent of $y_f$. Thus, we obtain
\begin{eqnarray} \label{box1}
& & \Xi_1 = \frac{1}{2} \left( \frac{M}{2} \right)^4 W_{\alpha}
W_{\beta}^2 W_{\gamma} \int dy_a dy_c dx_f dx_h \nonumber \\
& & \times \exp\left(-\frac{L_y-y_a}{l} - \frac{L_y-y_c}{l} -
\frac{x_f}{l} - \frac{L_x - x_h}{l} \right) \int d\bbox{r}_b \nonumber
\\
& & \times d\bbox{r}_d d\bbox{r}_e d\bbox{r}_g \tilde
P[X_{\beta},y_a;\bbox{r_b}] \tilde P[X_{\beta},y_c;\bbox{r_d}] \tilde
P[\bbox{r_e}; x_f, Y_{\alpha}] \nonumber \\
& & \times \tilde P[\bbox{r_g}; x_h, Y_{\gamma}] F(\bbox{r}_b,
\bbox{r}_e, \bbox{r}_d, \bbox{r}_g) 
\end{eqnarray}
and
\begin{eqnarray} \label{box2}
& & \Xi_3 = \frac{1}{2} \left( \frac{M}{2} \right)^4 W_{\alpha}
W_{\beta} W_{\gamma} W_{\delta} \int dy_a dy_c dx_f dx_h \nonumber \\
& & \times \exp\left(-\frac{L_y-y_a}{l} - \frac{y_c}{l} -
\frac{x_f}{l} - \frac{L_x - x_h}{l} \right) \int d\bbox{r}_b \nonumber
\\
& & \times d\bbox{r}_d d\bbox{r}_e d\bbox{r}_g \tilde
P[X_{\beta},y_a;\bbox{r_b}] \tilde P[X_{\delta},y_c;\bbox{r_d}] \tilde
P[\bbox{r_e}; x_f, Y_{\alpha}] \nonumber \\
& & \times \tilde P[\bbox{r_g}; x_h, Y_{\gamma}] F(\bbox{r}_b,
\bbox{r}_e, \bbox{r}_d, \bbox{r}_g). 
\end{eqnarray}
Here $Y_{\alpha}$, $X_{\beta}$, $Y_{\gamma}$, and $X_{\delta}$ denote
the positions of the corresponding leads. 

We see already from Eqs. (\ref{box1}) and (\ref{box2}) that the
results are not-universal in the sense that they depend on the
geometry of the sample. Indeed, within the approximation in which we
replace $P$ by $\tilde P$, the quantity $\Xi_1$ does not contain
any information on the location and width of lead $\delta$; at the
same time, it depends essentially on the location and width of other
leads. The quantity $\Xi_2$ contains information of all leads
except $\beta$, whereas both $\Xi_3$ and $\Xi_4$ are governed by
the geometry of all leads. Therefore all ratios $\Xi_i/\Xi_j$ depend
essentially on the geometry of the sample. This is in contrast with
the case of a chaotic cavity \cite{Langen}, where one obtains $\Xi_1 =
\Xi_2 = -3 \Xi_3 = -3 \Xi_4$ irrespectively of geometry, provided the
leads are wide enough.  

Performing the integration and taking into account that the remaining
sums are converging rapidly for $L_x \sim L_y$ (the case we assume
from now on), one obtains cumbersome expressions for the quantities
$\Xi_i$. In the symmetric case, $L_x = L_y = L$, $W_{\lambda} = W$,
$Y_{\alpha} = Y_{\gamma} = X_{\beta} = X_{\delta} = L/2$ they
simplify. We obtain 
\begin{eqnarray} \label{boxf}
\left\{ \begin{array}{c} 
\Xi_1 = \Xi_2 \\ \Xi_3 = \Xi_4 \end{array} \right\}
= \left\{ \begin{array}{c} \eta_1 \\ -\eta_3 \end{array} \right\}
p_Fl \left( \frac{W}{L} \right)^4,
\end{eqnarray}
with positive constants 
$$\eta_1 = \frac{1}{2\sinh^3 \pi}(\cosh \pi - 1)(2\pi\cosh \pi - \sinh
\pi) \approx 0.21\ ,$$
and
$$\eta_3 = \frac{1}{\sinh^3 \pi}(2\pi\cosh \pi - \sinh\pi) \approx
0.03\ .$$
It is seen that the exchange effect exists, and has a {\em negative}
sign (i.e. exchange suppresses the result of experiment C in comparison
with the sum of the results of experiments A and B). Although the
relative value of the effect is $\Xi_3/\Xi_1 \sim 0.1$, the effect
should be clearly observable. 

\subsection{Cross geometry}

We consider now the cross geometry of Fig.~1b. We assume that all arms
of the cross have equal \cite{foot} lengths $L$ and widths $W$. For $L
\gg W$ we can consider diffusion as one-dimensional. We also assume
that the center of the cross is described by a reflection coefficient
$R$ and a transmission coefficient $T =(1-R)/3$ between any two
different arms.   

The diffusion propagator is a solution of Eq. (\ref{diff2}). We move
to the coordinate system of Fig.~1b and fix the point $\bbox{r}$ near
the origin of the lead $\alpha$, $x \simeq L$. We introduce
$$P_{\alpha \lambda} (x, x') = P(x,x'),\ \ \ \mbox{if $x'$ lies in the
arm $\lambda$},$$
which is proportional to the time-integrated probability of diffusion
from point $x$ in the arm $\alpha$ to point $x'$ in the arm $\lambda$. 
The solution satisfying the boundary conditions and the condition of
current conservation in the cross, 
$$\sum_{\lambda} \partial_{x'} P_{\alpha\lambda} (x, x')\vert_{x' =
0} = 0, $$ 
is
\begin{eqnarray} \label{cross1}
P_{\alpha\alpha} (x,x_{\alpha}) & = & \frac{1}{MD\tau^2W}
\frac{(L-x)(L\epsilon + 3x_{\alpha})}{3+\epsilon}, \ \ \ x >
x_{\alpha} \nonumber \\ 
P_{\alpha\lambda} (x,x_{\lambda}) & = & \frac{1}{MD\tau^2W}
\frac{(L-x)(L - x_{\lambda})}{3+\epsilon}, \ \ \ \lambda \ne \alpha.
\end{eqnarray}
The constant $\epsilon$, defined as the ratio of diffusion
probabilities,  
\begin{equation} \label{cons}
\epsilon = \frac{P_{\alpha\alpha} (x,0)}{P_{\alpha\beta} (x,0)},
\end{equation}
is calculated in the Appendix. The result is 
\begin{eqnarray} \label{eps}
\epsilon = \left\{ \begin{array}{lr}
1 + l(LT)^{-1}(1-2T), & \ T \gg l/L \\
l(LT)^{-1}, & \ T \ll l/L  \end{array} \right. .
\end{eqnarray}

Now we substitute Eq. (\ref{cross1}) into the
general expressions (\ref{xi1g}) and (\ref{xi3g}). Since the area of
the cross is negligible in comparison with the areas of the arms, we
can neglect the possibility of finding the Hikami box inside the cross,
and allow it to be situated only in one of the arms. Upon integration
we obtain 
\begin{eqnarray} \label{cross2}
\Xi_1 & = & \Xi_2 = \frac{l}{3L}Wp_F \frac{3(1 + \epsilon^2) +
4}{(3+\epsilon)^4}, \nonumber \\ 
\Xi_3 & = & \Xi_4 = \frac{4l}{L} Wp_F \frac{\epsilon -
1}{(3+\epsilon)^4}. 
\end{eqnarray} 

Thus, in the case $T \gg l/L$, when the overall transmission through
the sample is governed by the diffusive arms rather than by the center
of the cross, one has $\epsilon \sim 1$. The quantities $\Xi_1$ and
$\Xi_2$ are regular for $\epsilon = 1$, and therefore assume the
finite value, $\Xi_1 = \Xi_2 = (5/192)(p_FWl/L)$. At the same time,
the exchange terms $\Xi_3$ and $\Xi_4$ are strongly suppressed in the
parameter $l/L$, $\Xi_3 = \Xi_4 = (1/64)(p_FWl^2/L^2T)(1-2T)$. In the
less realistic case $T \ll l/L$ (the transmission is determined by the
center of the cross) one obtains $\epsilon \gg 1$. All quantities
$\Xi_1$ are small, since now all channels are nearly closed (cf. the
situation for two-terminal shot noise \cite{B90,Khlus}), however
exchange terms are additionally suppressed in the parameter
$\epsilon^{-1}$.   

Thus, in the cross geometry of Fig.~1b the exchange noise $\langle
\Delta S \rangle$ is suppressed in comparison with the regular terms
$\langle S_A + S_B \rangle$ irrespectively of the transmission
properties of the center of the cross. It is also quite remarkable
that for the cross geometry the exchange contribution is {\em
positive}, although small: the total effect is enhanced by the exchange.   

\section{Conclusions}

We have investigated shot noise in diffusive conductors on the basis
of Eq. (\ref{HBT}) and the Fisher-Lee relation, which expresses
scattering matrices through advanced and retarded Green's
functions. In this way, one can reduce disorder averages of various
combinations of scattering matrices to standard diagram technique for
Green's functions \cite{AA}. Although this approach resembles
previously published calculations of conductance and conductance
fluctuations \cite{IWZ,Altland,IM,MG,MMZ}, we believe it to be more
transparent. We are not aware of any applications of this approach to
noise problems. 

As a check of the method, we first reproduced the $1/3$-shot noise
suppression in the two-terminal geometry and confirmed the statement
of Ref. \cite{Nazarov} that it is in fact super-universal and holds for
an arbitrary relation between length and width of a wire, provided the
system is diffusive. Our proof bears some similarity with other ones
existing in the literature \cite{BB,Nagaev,ALY}; however, it is
novel, and a direct equivalence to any of the existing proofs
is not evident (see the discussion in the end of Section 2).  

Then we turned to the multi-terminal geometry and investigated the
interference experiment, similar to the Hanbury Brown and Twiss
experiment known in optics \cite{optics}. We obtained general
expressions for scattering matrix combinations (\ref{xi1g}),
(\ref{xi3g}), determining noise intensities (\ref{basic}); then we
investigated them for the two different geometries of Fig.~1.
\begin{figure}
\centerline{\psfig{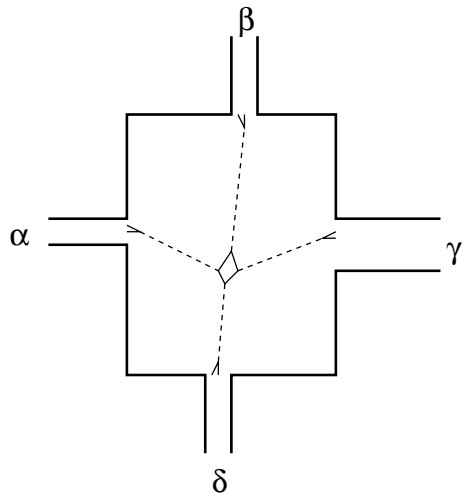}}
\vspace{0.3cm}
\caption{Typical electron trajectories, contributing to the quantity 
$\Xi_3$. Solid lines denote ballistic propagation (described by
averaged single-particle Green's function), and dashed lines denote
diffusive propagation (described by the diffuson $P$).}   
\label{gdiag6}
\end{figure}

The important point we make is that the exchange effect, even when
averaged over disorder, does not vanish. The reason is that typical
electron trajectories, contributing to {\em all} averaged traces of
scattering matrices, considered above (i.e. quantities $g$ and $t$ for
the two-terminal geometry, and $\Xi_i$ in the four-terminal case) do
not contain large closed loops. In particular, it is valid for the
``exchange'' traces $\Xi_3$ and $\Xi_4$. A typical trajectory for the
quantity $\Xi_3$ is shown in Fig.~6. It is a direct translation of
diagrams contributing to this quantity. The electron motion is
essentially diffusion between different leads with ballistic
propagation (described by disorder-averaged single-particle Green's
function) close to the leads and somewhere in the middle of the sample
(the later motion described by the Hikami box in
Eq. (\ref{xi3g})). Thus, closed loops are related to ballistic motion
over distances of an elastic scattering length only, and therefore
neither the shot noise in two-terminal conductors nor the shot noise
in multi-terminal structures should be sensitive to dephasing.   

Another observation is that exchange corrections are not
universal, in contrast to what is found in the chaotic case
\cite{Langen}: the ratio $\langle \Delta S \rangle/\langle S_A + S_B
\rangle$ depends on the geometry of a sample in an essential way. Even
the sign of the effect may change: for the box geometry of Fig.~1a it
is negative, i.e. interference suppresses the total effect, while for
the cross geometry (Fig.~1b) interference enhances the effect
(although weakly).  

The results obtained for the cross geometry allow us to make
predictions for experiments in real systems. Indeed, we found that the
exchange contribution is suppressed strongly with respect to the
average noise intensities $\langle S_A \rangle$ and $\langle S_B
\rangle$. This result was obtained by assuming that the intermediate
scattering, described by the Hikami box, does not happen in the center
of the cross, i.e., strictly speaking, for ballistic propagation
through the center. In more complicated situations the entire exchange
effect will be determined by properties of the center of the cross. If
the motion within the center is diffusive, one can apply the results
obtained above for the box geometry. The total exchange effect is
expected to be negative. However, since the arms of the cross (which
correspond to disordered leads in the real experiments) contribute to
the intensities $\langle S_A \rangle$ and $\langle S_B \rangle$, but not
to the exchange contribution, the latter will still be suppressed, if
disorder extends far into leads. Finally, if the center of the cross
is a chaotic cavity, one may use the results of
Ref. \cite{Langen}. The exchange contribution in the chaotic cavity
separated from ideal leads by high barriers (disordered arms play the
role of these barriers) is positive: the interference enhances the
effect.      

\section*{Acknowledgments}

We thank S.~van Langen, who calculated the exchange-interference
correlator for chaotic cavities, for useful discussions. The work was
supported by the Swiss National Science Foundation.  

\section*{Appendix}

To find the coefficient $\epsilon$ defined by Eq. (\ref{cons}) it is
instructive to consider a discrete model of diffusion \cite{B87}. Each
arm is modeled by a one-dimensional array of scatterers, placed at
a distance $l$ from each other; the total number of scatterers in each
arm is $N = L/l$. Each scatterer is described by transmission $t=1/2$ and
reflection $r=1/2$ probabilities. We denote the carrier flux densities
in the arm $\alpha$ between sites $n$ and $n+1$ away from the center
of the cross by $a_n$, and the flux towards the center of the cross by
$b_n$. Corresponding amplitudes in other arms are denoted by $a'_n$
and $b'_n$ (Fig.~7). The total flux at each site is given by $\rho_n =
a_n + b_n$, and $\rho'_n = a'_n + b'_n$. The coefficient $\epsilon$
can be expressed as $\epsilon = \rho_0/\rho'_0$.  
\begin{figure}
\centerline{\psfig{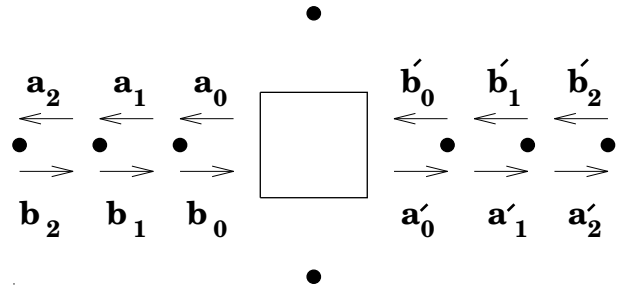}}
\vspace{0.3cm}
\caption{Discrete diffusion model.}   
\label{gdiag7}
\end{figure}
The diffusion equation implies that all densities should be linear
functions of $n$; furthermore, matching conditions at each scatterer
require $b_{n-1} = a_{n}$; $b'_{n-1} = a'_n$. Thus, we write
$$a_n = A + B(n-1), \ \ a'_{n} = A' + B'(n-1), \eqno(A1)$$ 
$$b_n = A + Bn, \ \ b'_{n} = A' + B'n. \eqno(A2)$$
The four constants $A, B, A', B'$ obey four equations:

1) Boundary condition for the arm $\beta$: $b'_N = 0$. 

2) Matching conditions at the center of the cross:
$$a'_0 = Tb_0 + 2Tb'_0 + Rb_0, \eqno(A3)$$

and
$$a_0 = 3Tb'_0 + Rb_0. \eqno(A4)$$

The fourth equation is Eq.(\ref{diff2}), however, it is not
required for the calculation of the constant $\epsilon$. We obtain 
$$\epsilon = \frac{2(N + T^{-1}) - 3}{2N+1}, \eqno(A5)$$
and the limiting cases given by Eq. (\ref{eps}) follow immediately.


\begin{thebibliography}{99}

\bibitem{BdJ} For a review of shot noise in electron
systems, see C.~W.~J.~Beenakker and M.~J.~M.~de~Jong,
cond-mat/9611140, to be published in: {\em Mesoscopic Electron
Transport}, ed. by L.~P.~Kouwenhoven, G.~Sch\"on, and L.~L.~Sohn, NATO
ASI Series E (Kluwer Academic Publishing, Dordrecht).  

\bibitem{BB} C.~W.~J.~Beenakker and M.~B\"uttiker, Phys. Rev. B {\bf
46}, 1889 (1992). 

\bibitem{Nagaev} K.~E.~Nagaev, Phys. Lett. {\bf A 169}, 103 (1992). 

\bibitem{ALY} B.~L.~Altshuler, L.~S.~Levitov, and A.~Yu.~Yakovets,
Pis'ma Zh. Eksp. Teor. Fiz. {\bf 59}, 821 (1994) [JETP Lett. {\bf 59},
857 (1994)].

\bibitem{Nazarov} Yu.~V.~Nazarov, Phys. Rev. Lett. {\bf 73}, 134 (1994).

\bibitem{Liefrink} F.~Liefrink, J.~I.~Dijkhuis, M.~J.~M.~de Jong,
L.~W.~Molenkamp, and H.~van Houten, Phys. Rev. B {\bf 49}, 14066
(1994). 

\bibitem{semicl} M.~J.~M.~de~Jong and C.~W.~J.~Beenakker, Phys. Rev. B
{\bf 51}, 16867 (1995); Physica {\bf A 230}, 219 (1996). 

\bibitem{Steinbach} A.~H.~Steinbach, J.~M.~Martinis, and
M.~H.~Devoret, Phys. Rev. Lett. {\bf 76}, 3806 (1996). 

\bibitem{Nagaev2} K.~E.~Nagaev, Phys. Rev. B {\bf 52}, 4740 (1995). 

\bibitem{Kozub} V.~I.~Kozub and A.~M.~Rudin, Phys. Rev. B {\bf 52},
7853 (1995). 

\bibitem{Shimizu} A.~Shimizu and M.~Ueda, Phys. Rev. Lett. {\bf 69},
1403 (1992).

\bibitem{Landauer} R.~Landauer, Ann. N.~Y. Acad. Sci. {\bf 755}, 417
(1995); Physica {\bf B 227}, 156 (1996).  

\bibitem{Liu} R.~C.~Liu and Y.~Yamamoto, Phys. Rev. B {\bf 50}, 17411
(1994); {\em ibid} {\bf 53}, 7555(E) (1994). 

\bibitem{B92} M.~B\"uttiker, Phys. Rev. B {\bf 46}, 12485 (1992). 

\bibitem{optics} R.~Hanbury Brown and R.~Q.~Twiss, Nature {\bf 177}, 27
(1956); M.~L.~Goldberger, H.~W.~Lewis, and K.~M.~Watson,
Phys. Rev. {\bf 132}, 2764 (1963); R.~Loudon, in: {\em Disorder in
Condensed Matter Physics}, Ed. by J.~A.~Blackman and J.~Taguena
(Clarendon Press, Oxford, 1991), p.441.

\bibitem{foot1} We set $\omega = T = 0$ and discuss only the regime
linear in voltage $V$ throughout the paper. We also set $\hbar = 1$.  

\bibitem{foot2} Current correlations in normal conductors are quite
generally negative. Thus, $S_j$ as defined here are positive
quantities in normal conductors. In contrast, in hybrid normal and
superconducting structures the current-current correlations can change
sign. See M.~P.~Anantram and S.~Datta, Phys. Rev. B {\bf 53}, 16390
(1996); T. Martin, Phys. Lett. {\bf A 220}, 137 (1996). 

\bibitem{Langen} S.~van~Langen and M.~B\"uttiker (unpublished).

\bibitem{Stern} F.~von Oppen and A.~Stern, cond-mat/9611079.

\bibitem{FLee} D.~S.~Fisher and P.~A.~Lee, Phys. Rev. B {\bf 23}, 6851
(1981).

\bibitem{Stone} A.~D.~Stone and A.~Szafer, IBM J. Res. Develop. {\bf
32}, 384 (1988). 

\bibitem{BS} H.~U.~Baranger and A.~D.~Stone, Phys. Rev. B {\bf 40},
8169 (1989).

\bibitem{AGD} A.~A.~Abrikosov, L.~P.~Gor'kov, and
I.~E.~Dzyaloshinski, {\em Methods of Quantum Field Theory in
Statistical Physics} (Prentice-Hall, Englewood Cliffs, N.J. 1963). 

\bibitem{IWZ} S.~Iida, H.~A.~Weidenm\"uller, and J.~A.~Zuk,
Ann. Phys. (N.Y.) {\bf 200}, 219 (1990).

\bibitem{Altland} A.~Altland, Z. Phys. {\bf B 82}, 105 (1991).

\bibitem{IM} S.~Iida and A.~M\"uller-Groeling, Phys. Rev. B {\bf 44},
8097 (1991). 

\bibitem{MG} A.~M\"uller-Groeling, Phys. Rev. B {\bf 47}, 6480 (1993).   

\bibitem{MMZ} A.~D.~Mirlin, A.~M\"uller-Groeling, and M.~R.~Zirnbauer,
Ann. Phys. (N.Y.) {\bf 236}, 325 (1994).

\bibitem{B90} M.~B\"uttiker, Phys. Rev. Lett. {\bf 65}, 2901 (1990).

\bibitem{Khlus} V.~A.~Khlus, Zh. Eksp. Teor. Fiz. {\bf 93}, 2179
(1987) [Sov. Phys. -- JETP {\bf 66}, 1243 (1987)]; 
G.~B.~Lesovik, Pis'ma Zh. Eksp. Teor. Fiz {\bf 49}, 513 (1989) [JETP
Lett. {\bf 49}, 592 (1989)].

\bibitem{AA} B.~L.~Altshuler and A.~G.~Aronov, in: {\em
Electron-electron Interactions in Disordered Systems}, Ed. by
A.~L.~Efros and M.~Pollak (North-Holland, Amsterdam, 1985), p.1.

\bibitem{Hikami} S.~Hikami, Phys. Rev. B {\bf 24}, 2671 (1981).

\bibitem{KhL} This derivation is close to the calculation of
fluctuations of current-voltage characteristics in mesoscopic
conductors, A.~I.~Larkin and D.~E.~Khmel'nitskii,
Zh. Eksp. Teor. Fiz. {\bf 91}, 1815 (1986) [Sov. Phys. -- JETP, {\bf
64}, 1075 (1986)]. 

\bibitem{DMPK} O.~N.~Dorokhov, Pis'ma Zh. Eksp. Teor. Fiz {\bf 36},
259 (1982) [JETP Lett. {\bf 36}, 318 (1982)]; P.~A.~Mello, P.~Pereyra,
and N.~Kumar, Ann. Phys. (N.Y.) {\bf 181}, 290 (1988).

\bibitem{foot} Generalization to arbitrary geometry is trivial
provided $L \gg W$, and does not lead to any qualitatively new
results. 

\bibitem{B87} A similar approach was used to investigate the time of
diffusion by R.~Landauer and M.~B\"uttiker, Phys. Rev. B {\bf 36}, 6255
(1987), and for the description of the dephasing electrode by
M.~B\"uttiker, {\em ibid}, {\bf 35}, 4123 (1987). 

\end{thebibliography}
\end{document}